\documentclass[12pt, letterpaper]{article}
\usepackage[top=1in, bottom=1.5in, left=1in, right=1in]{geometry}
\usepackage[utf8]{inputenc}
\usepackage{booktabs}
\usepackage{hyperref}

\usepackage{graphicx}
\usepackage{longtable}

\newcommand{\degr}{\ensuremath{^\circ}}

\title{Simultaneous LSST and Euclid observations -- advantages for Solar System Objects}
\author{C. Snodgrass, B. Carry, J. Berthier, S. Eggl, M. Mommert, \\J.-M. Petit, F. Spoto, M. Granvik, R. Laureijs, B. Altieri, \\
R. Vavrek, L. Conversi, A. Nucita, M. Popescu, G. Verdoes Kleijn, \\
M. Kidger, G. H. Jones, D. Oszkiewicz, M. Juric, L. Jones\\
for the \textsl{Euclid Solar System Object Science Working Group}\\
(SSO SWG) }
\date{Nov 2018}

\begin{document}

\maketitle

\begin{abstract}
The ESA Euclid mission is a space telescope that will survey $\sim$15,000 square degrees of the sky, primarily to study the distant universe (constraining cosmological parameters through the lensing of galaxies). It is also expected to observe $\sim$150,000 Solar System Objects (SSOs), primarily in poorly understood high inclination populations, as it will mostly  avoid $\pm$15\degr~from the ecliptic plane. With a launch date of 2022 and a 6 year survey, Euclid and LSST will operate at the same time, and have complementary capabilities.  We propose a LSST mini-survey to coordinate quasi-simultaneous observations between these two powerful observatories, when possible, with the primary aim of greatly improving the orbits of SSOs discovered by these facilities. As Euclid will operate from a halo orbit around the Sun-Earth L2 Lagrangian point, there will be significant parallax between observations from Earth and Euclid (0.01 AU). This means that simultaneous observations will give an independent distance measurement to SSOs, giving additional constraints on orbits  compared to single Euclid visits.
\end{abstract}

\section{White Paper Information}
Corresponding authors: Colin Snodgrass (csn@roe.ac.uk), 
Benoit Carry (benoit.carry@oca.eu)
\begin{enumerate} 
\item {\bf Science Category:} Taking an Inventory of the Solar System
\item {\bf Survey Type Category:} Other: suggested scheduling to enable quasi-simultaneous observations with another survey telescope 
\item {\bf Observing Strategy Category:} 
an integrated program with science that hinges on the combination of pointing and detailed 
	observing strategy.
\end{enumerate}

\clearpage

\section{Scientific Motivation}



\begin{figure}
     \centering
     \includegraphics[width=0.8\hsize]{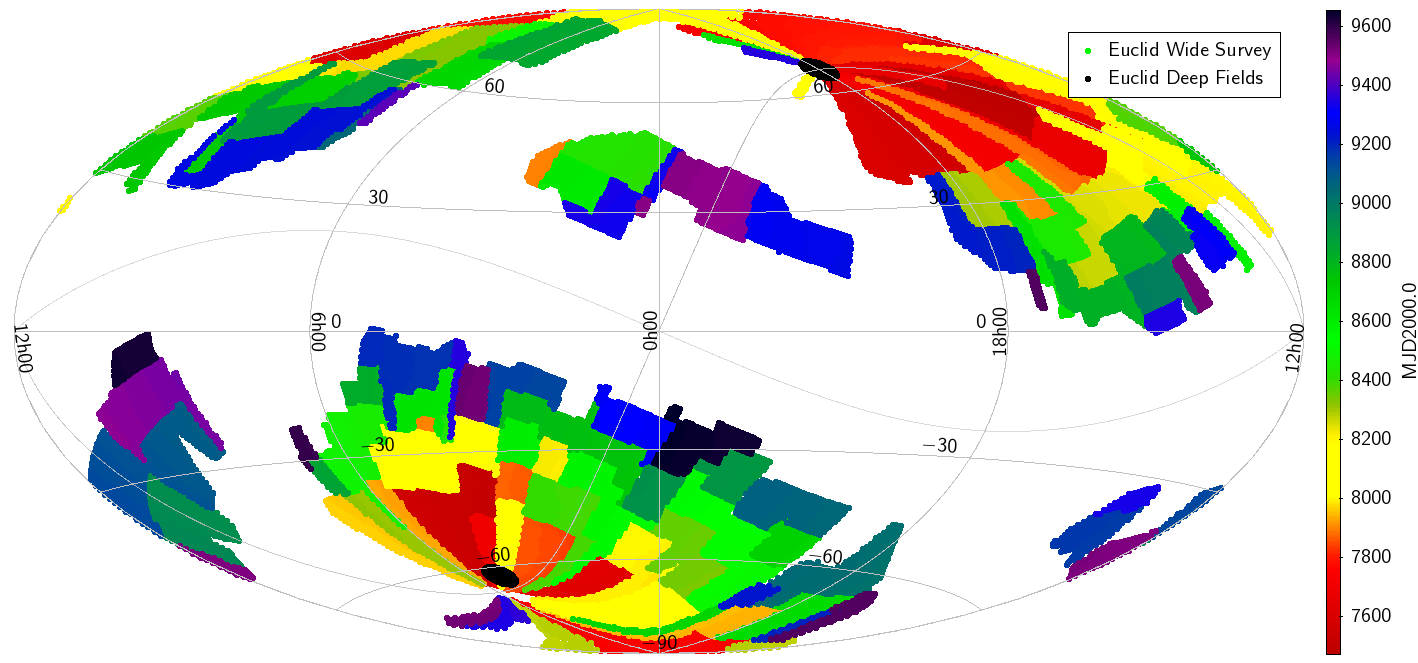}
     \caption{Expected coverage of the
Euclid Survey color-coded by observing epoch, in equatorial coordinates. The grey line represents the ecliptic plane. The two circular black regions are the location of the Euclid Deep Fields North and South. 
}
     \label{fig:rs}
 \end{figure}

Euclid is the European Space Agency (ESA)'s mission to ``map the geometry of the dark Universe'', improving our understanding of dark matter and dark energy, by studying the distortion of distant galaxies due to weak gravitational lensing. Euclid is designed to provide diffraction-limited imaging in both the visible and near-IR, with a 1.2m diameter telescope. It is a survey mission, designed to image a wide area of the sky (15,000 deg$^2$), but, apart from some Deep Survey fields covering 40 deg$^2$ and some calibration fields, will visit each field only once. However, each $\sim$hour-long visit is composed of 4 blocks, each comprising a 565s image in the broad visible band and shorter exposures in Y (121s), J (116s), and H (81s) plus slitless spectroscopy in near-IR. Solar System Objects (SSOs) will generally appear as trailed objects (depending on their relative speed, with nearby objects most trailed) and will be recognizable by their motion between the four consecutive imaging blocks. Although Euclid will avoid ecliptic latitudes within 15\degr~of the ecliptic plane (Fig.~\ref{fig:rs}) to avoid  zodiacal dust emission, its sensitivity to faint objects (detection limit of $m_{AB} = 24.5$ for 10$\sigma$ on a 1" extended source) and wide field of view ($0.8 \times 0.7$ deg$^2$) means that it will still detect large numbers of SSOs, and the Euclid consortium now includes a SSO working group to drive the exploitation of this data set.

The Euclid survey strategy is fixed well in advance (any change in the schedule will be decided at least 6 weeks prior to the observations), and allows simulation of where Euclid will be pointing at a particular time, and therefore what known SSOs will be in the fields. Carry (2018) ran this simulation, also including estimated populations of fainter unknown objects to test Euclid's discovery potential, finding that Euclid can expect to observe around 150,000 SSOs (only about 1\% of which are currently known objects).



\begin{figure}
    \centering
    \includegraphics[width=0.8\hsize]{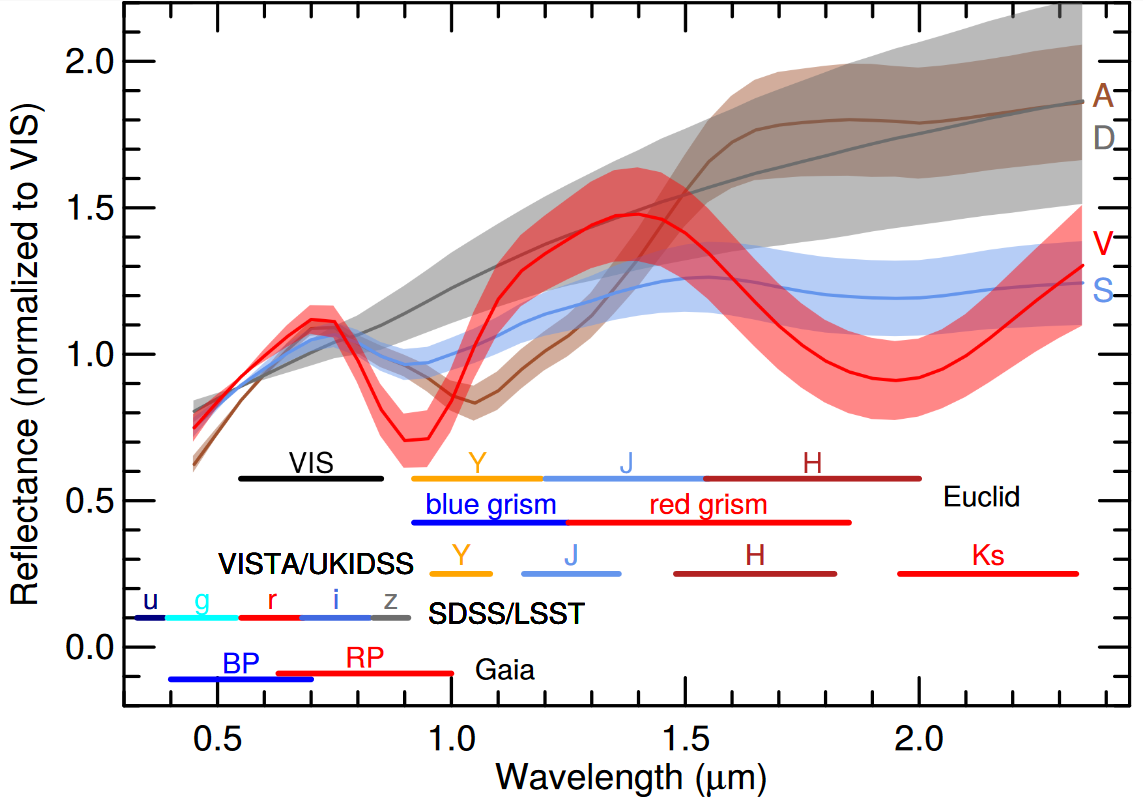}
    \caption{Spectral degeneration of asteroid spectral classes over visible only (including LSST filters). Euclid near-infrared will disentangle these classes.}
    \label{fig:degen}
\end{figure}

Euclid provides spectroscopic information and near-IR colours that are diagnostic of different mineral features, breaking the degeneracy that exists between several spectral classes of asteroids at visible wavelengths  (Fig.~\ref{fig:degen}). Euclid can therefore map compositional variation in faint objects better than a visible only survey, naturally supplementing LSST's capabilities. Finally, the spatial resolution offered by a space telescope (0.1" pixels in the VIS band) makes it sensitive to any binarity or extended features (i.e., cometary activity) around SSOs. Probing the  cometary activity of Centaurs will be a particular strength of Euclid, with the potential to constrain orbit evolution (planetary migration) models.

 LSST is complimentary to Euclid in time domain astronomy: LSST is bound to discover many more objects as it covers the ecliptic plane, and because it returns to fields many times, meaning that the orbits of SSOs can be determined. Orbits from the short arc of Euclid data alone will be very uncertain, and completely unconstrained in many cases. LSST observations will also allow sparse photometry (i.e., with a typical separation in time larger than the rotation period) lightcurves to be built up over the length of the surveys, to study rotation and shape of SSOs, their photometric phase curves, and will allow visible colours to be measured (Euclid has only a single broad visible passband).


Euclid is expected to launch in 2022 and perform a 6-year survey, meaning that parallel observations between LSST and Euclid should be possible for much of the time. Considering the comparable photometric depth 
of both surveys, there is therefore the possibility to combine their strengths to obtain more detailed information on the SSOs seen by both (100s per field).
Some of this can be done {\it post facto} by using LSST determined orbits to link with observations obtained by Euclid, 
but in this white paper we argue that optimizing a relatively small amount of the LSST survey to obtain \emph{simultaneous} observations between LSST and Euclid, when possible, has significant advantages. 

LSST and Euclid have comparable imaging depths and sensitivities. 
For orbit determination, it presents the unique opportunity to constrain distances to the observed objects via parallaxes, as Euclid will be observing from the Sun-Earth L2 Lagrangian point 1.5 million km from Earth. 
Two sets of astrometric positions simultaneously observed by Euclid and LSST could allow for preliminary orbits of solar system objects that may otherwise be lost (e.g. Eggl 2011).
 While many Euclid SSOs will be eventually observed by LSST anyway, and their orbits determined independently, only parallel observations will allow rapid determination of  orbits for recovery of  rare objects, allowing follow up elsewhere, and direct comparison between visible and near-IR observations at the same geometry.
For photometry, simultaneous observations would remove systematic lightcurve effects due to the rotation of irregular (unknown) shapes that otherwise drastically limit the precision of SSO colors.
For nearby objects, the difference in viewing geometry may also allow some constraints to be placed on phase angle effects on photometry.
Secondly, the addition of LSST visible and Euclid near-IR photometry will improve compositional characterisation.


LSST is the only realistic facility to perform these observations, reaching a similar depth to Euclid observations and having a wide field of view. Whether or not Euclid will provide its own  alert stream for SSO discoveries is still under discussion, but in the case that it does, significant 8m telescope time would be required to confirm orbits for interesting objects. 
Finally, we also expect that other (non SSO) communities may find some advantage in simultaneous LSST+Euclid observations, especially given the near-IR spectroscopic coverage of Euclid, e.g., transient events.
 
\vspace{.6in}


\clearpage

\section{Technical Description}

\subsection{High-level description}

{\bf We propose small modifications to the LSST exposure scheduling during a fraction of the survey so that, when possible, Euclid fields are observed simultaneously by the spacecraft and LSST.} This will be a minimal fraction of total LSST pointings, and we propose that beyond this the survey continues as planned, with repeats of the Euclid fields once more within the night and then again as another pair within a few days to one month (depending on what the final repeat cadence is set as for the main survey). We do not require any special repeat pattern beyond that required for LSST MOPS to recognise SSOs. The observations of the Euclid fields should be to standard LSST depth, and we have only weak constraints on the filter choice (should avoid the less sensitive filters to achieve a good single visit depth).

\begin{figure}
    \centering
    \includegraphics[width=\hsize]{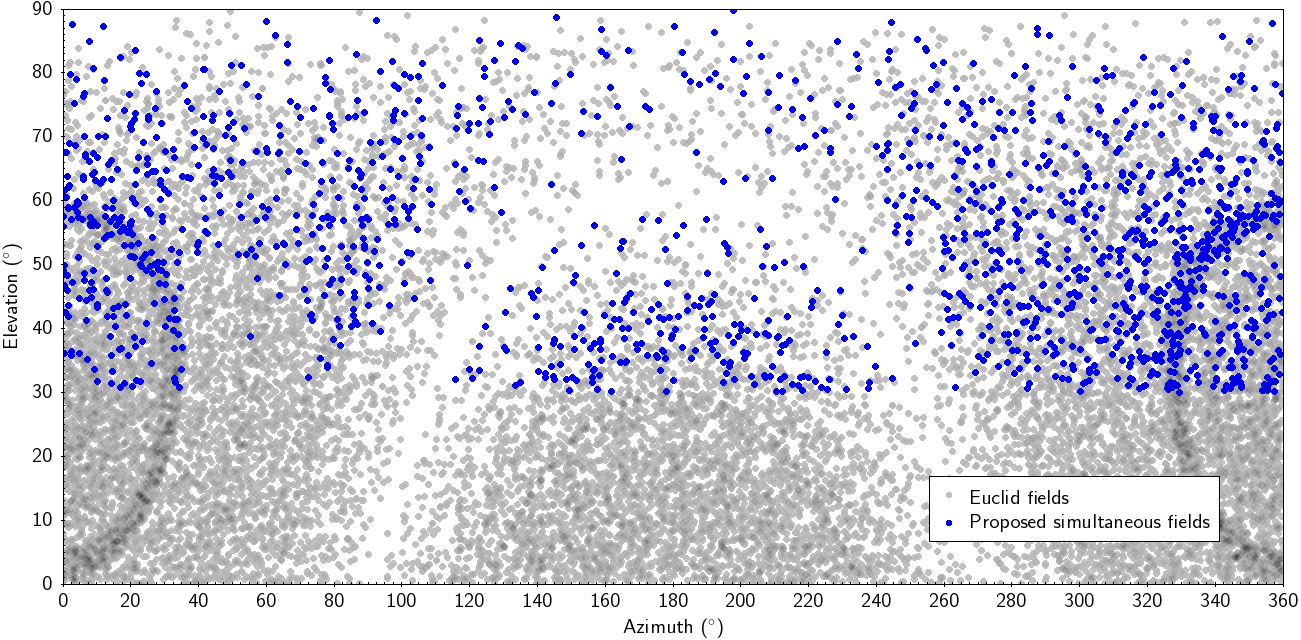}
    \caption{Distribution of Euclid fields in LSST local frame (azimuth-elevation), with the proposed simultaneous field highlighted (blue dots).}
    \label{fig:azh}
\end{figure}

\begin{figure}
    \centering
    \includegraphics[width=\hsize]{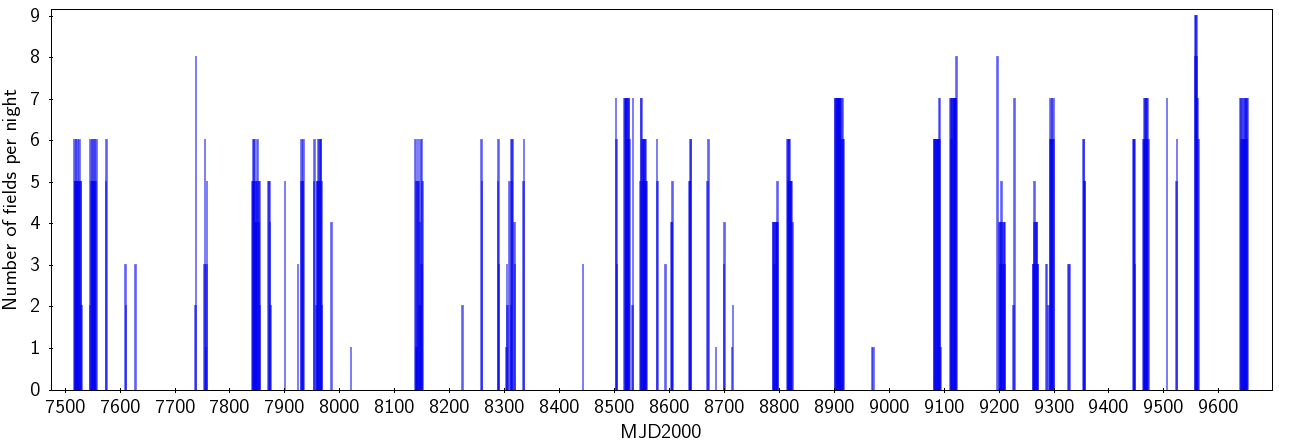}
    \caption{Distribution in time of the proposed fields, corresponding typically to a handful during a night.}
    \label{fig:time}
\end{figure}

Around 8\% of Euclid fields will be observable (see below) from LSST during the night time, or 3,489 pointings over 6 years (Fig.~\ref{fig:azh}). As the planned single visit depth of LSST in the standard Wide-Fast-Deep survey is comparable with Euclid (for observations in $g$, $r$, or $i$), \textbf{we do not suggest extra observations are required to perform parallel observations, but these would place some constraints on observation timing (and therefore cadence)}, but given the expected $\sim$1000 LSST pointings per night, the impact of this scheduling should be minimal  (at most a handful pointings during a given night, see Fig.\ref{fig:time}).

The nominal Euclid survey fields, and the timing of these observations, is already known. Although this is likely to be modified based on the eventual launch date of the spacecraft (some delays are always probable in launching space telescopes), once fixed the timing of individual observations will be known years in advance. In the case of any change in the planning of Euclid survey, the decision will be known at least 6 weeks before the observations. This allows this information to be fed into LSST scheduling early; the proposed timing modifications are not going to be late-time ToO changes.

Based on the current baseline Euclid survey,  
we have analysed the Euclid pointing profile for visibility from LSST, and find that $\sim$8\% of Euclid fields are observable during the night from Cerro Pachon. 3,489 of the $\sim$42,000 Euclid pointings are observable at night (Sun below -18\degr~from the horizon) at elevation $> 30^\circ$, and about half of these are observable with both Sun and Moon distances  $> 90^\circ$, for more sensitive dark skies. Additionally, around 280 fields are observable (with the same airmass\,$<$\,2 limit) during each of the three twilights (civil, nautical, astronomical). Data taken in astronomical twilight and slightly closer to the Moon will still be useful, even if the fainter objects seen by Euclid are not detected.

\begin{figure}
    \centering
    \includegraphics[width=\hsize]{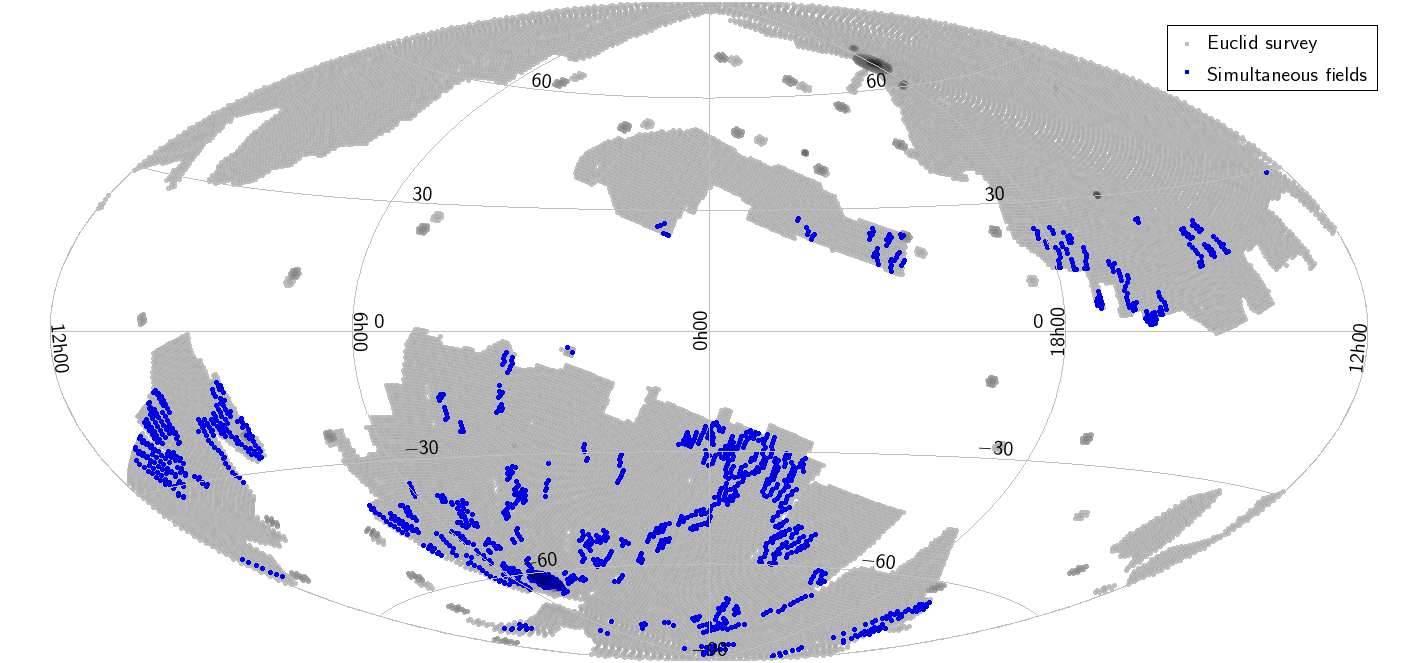}
    \caption{Distribution of fields (blue) eligible for simultaneous observations overprint on Euclid survey (grey) in RA/Dec projection. All fields have $\delta < 30\degr$.}
    \label{fig:simultaneous}
\end{figure}

Many of the observable fields correspond to the Southern spot (Euclid Deep Field South) which will be visited many times (another Deep Field, corresponding to the Chandra deep field in Fornax constellation, will be observed by Euclid but was not included in present computation as the survey simulation have yet to be produced), but others are spread across the sky from the wide survey (Fig.~\ref{fig:simultaneous}). As SSO detections will obviously increase toward the ecliptic, there is a clear bias against inclined population in current census (at least of asteroids, see Mahlke et al. 2018). Hence, any objects discovered by Euclid/LSST in these southern fields will be of high importance.
Based on these considerations, we find that there are 1,536 pointings (4\% of the Euclid total) that correspond to the SSO ``goldmine'':

\begin{itemize}
\item above elevation of 30\degr~from LSST
\item during the night (sun below -18\degr~elevation)
\item 90\degr~apart from both the moon and the sun
\end{itemize}

We believe that at least this set, corresponding to an almost negligible number of pointings for LSST over 6 years, should be high priority for simultaneous observation.
 

\vspace{.3in}

\subsection{Footprint -- pointings, regions and/or constraints}

We require specific pointings at specific times.
The timing constraint is however mild on LSST operation, with an hour-long window to acquire the observation each time.
The operating mode of Euclid encompasses 16 images of the same field obtained within approximately one hour, which allows some leverage to LSST schedule to achieve simultaneity: the observation can be acquired at any time within the hour-long time slot.

The final list of coordinates will be fixed once the final Euclid scheduling is set, but based on the baseline survey, we provide the list as an appendix at the end of the document (Table~\ref{tab:pointing}).

The vast majority of these pointings fall within the main LSST wide-fast-deep survey region. Some (particularly at dec $>$ 0, or those visible in twilights) correspond to areas proposed in other mini-surveys. Parallel observations with Euclid appear to be compatible with the requested observations by these surveys, with only adjustment in absolute timing of the relevant visit being necessary to achieve both sets of goals. 

\subsection{Image quality}

No constraints.

\subsection{Individual image depth and/or sky brightness}

The depth of Euclid to point sources in the broad VIS band is $m_{AB} = 24.5$ for 10$\sigma$ on a 1" extended source, which is a good match to the expected single visit LSST depth in any of the more sensitive bands (g, r or i). We therefore do not need to modify exposure times from the LSST survey, our constraints are only on timing of the observations, and filter choice.

\subsection{Co-added image depth and/or total number of visits}

Co-added image depth is not relevant, we are looking at SSOs in individual visits.
We require one visit parallel to the Euclid observations, one more the same night, and another pair of visits within a month (i.e., usual survey cadence) for the LSST MOPS to identify moving objects and find orbits. The only change to the usual LSST survey requested here is a scheduling constraint on  one of these visits.

\subsection{Number of visits within a night}

Two visits are required for MOPS, although only one of these needs have fixed timing to match Euclid. In fact we will be able to identify moving objects in a single frame by comparison with the Euclid observations, but operations will be greatly simplified if the LSST detections are automatically recognised by MOPS.

\subsection{Distribution of visits over time}

Fixed timing of observations.

\subsection{Filter choice}

Any of g, r, or i are acceptable as either of those filters will provide additional spectrophotometric information supplementing the Euclid VIS, Y, J, and H filter bands, and have a similar sensitivity to Euclid in a single visit.

\subsection{Exposure constraints}

Usual survey exposure times (assumed to be 30s) are appropriate, shorter exposures should not be used. No maximum (objects will always be significantly less trailed than Euclid observations). 

\subsection{Other constraints}

n.a.

\subsection{Estimated time requirement}

The proposed schedule does not change the time required for the LSST survey: we only argue in favor of a prioritization of certain fields at specific times to be obtained synchronously to Euclid observations and benefit from the parallax between the two facilities.
Given Euclid operations, in which Euclid will pave the sky by acquiring images of the same field over an hour, the timing constraint for LSST observation will be mild: the LSST observation of the field can be obtained at any time within the hour-long slot.

\vspace{.3in}

\begin{table}[ht]
    \centering
    \begin{tabular}{l|l|l|l}
        \toprule
        Properties & Importance \hspace{.3in} \\
        \midrule
        Image quality &   3  \\
        Sky brightness & 2 \\
        Individual image depth &  1 \\
        Co-added image depth & 3  \\
        Number of exposures in a visit   & 3  \\
        Number of visits (in a night)  &  2 \\ 
        Total number of visits &  3 \\
        Time between visits (in a night) & 3 \\
        Time between visits (between nights)  &  3 \\
        Long-term gaps between visits & 3\\
        Other (absolute timing of visit) & 1\\
        \bottomrule
    \end{tabular}
    \caption{{\bf Constraint Rankings:} Summary of the relative importance of various survey strategy constraints. Please rank the importance of each of these considerations, from 1=very important, 2=somewhat important, 3=not important. If a given constraint depends on other parameters in the table, but these other parameters are not important in themselves, please only mark the final constraint as important. For example, individual image depth depends on image quality, sky brightness, and number of exposures in a visit; if your science depends on the individual image depth but not directly on the other parameters, individual image depth would be `1' and the other parameters could be marked as `3', giving us the most flexibility when determining the composition of a visit, for example.}
        \label{tab:obs_constraints}
\end{table}

\subsection{Technical trades}
{\it To aid in attempts to combine this proposed survey modification with others, please address the following questions:
\begin{enumerate}
    \item What is the effect of a trade-off between your requested survey footprint (area) and requested co-added depth or number of visits? {\bf n.a.}
    \item If not requesting a specific timing of visits, what is the effect of a trade-off between the uniformity of observations and the frequency of observations in time? e.g. a `rolling cadence' increases the frequency of visits during a short time period at the cost of fewer visits the rest of the time, making the overall sampling less uniform. {\bf n.a.}
    \item What is the effect of a trade-off on the exposure time and number of visits (e.g. increasing the individual image depth but decreasing the overall number of visits)? {\bf The number of visits is set by MOPS requirements to identify SSOs.}
    \item What is the effect of a trade-off between uniformity in number of visits and co-added depth? Is there any benefit to real-time exposure time optimization to obtain nearly constant single-visit limiting depth? {\bf n.a.}
    \item Are there any other potential trade-offs to consider when attempting to balance this proposal with others which may have similar but slightly different requests? {\bf n.a.}
\end{enumerate}}

\section{Performance Evaluation}

The simple metric that will describe the performance of any given survey strategy against our goals will be a percentage of the possible Euclid observations that are achieved within the necessary time windows (see table \ref{tab:pointing}). A minimum set will be the ``goldmine'' subset that are observable at larger solar and lunar elongation (see section 3.1).

\vspace{.6in}

\section{Special Data Processing}

No specific data processing beyond the MOPS processing to identify SSOs in the LSST data. Combining  simultaneous observations from both Euclid and LSST will allow distances to be derived by parallax -- these measurements will be made my team members with both LSST and Euclid data access rights using their own (non-project) software. 

\section{References}

Carry, B. (2018). Solar System Science with ESA Euclid. {\bf A\&A} 609, A113\\
\smallskip\\
Eggl, S. (2011). Refinement of Near Earth Asteroids? orbital elements via simultaneous measurements by two observers. {\bf Celestial Mechanics and Dynamical Astronomy}, 109(3), 211-228.\\
\smallskip\\
Mahlke, M., Bouy, H., Altieri, B., Verdoes Kleijn, G., Carry, B., Bertin, E., de Jong, J. T. A., Kuijken, K., McFarland, J. \& Valentijn, E. (2018).
Mining the Kilo-Degree Survey for solar system objects. {\bf Astronomy \& Astrophysics}, 610, A21.

\newpage

\begin{center}
\begin{longtable}{rrrrrrr}
\caption{Appendix: Preliminary list of pointings.} \label{tab:pointing} \\
\hline
  \multicolumn{1}{c}{MJD2000.0} &
  \multicolumn{1}{c}{RA} &
  \multicolumn{1}{c}{DEC} &
  \multicolumn{1}{c}{Az} &
  \multicolumn{1}{c}{Elevation} &
  \multicolumn{1}{c}{Distance to Sun} &
  \multicolumn{1}{c}{Distance to Moon} \\ \hline
  \endfirsthead

\multicolumn{7}{c}%
{{\bfseries \tablename\ \thetable{} -- continued from previous page}} \\
\hline
  \multicolumn{1}{c}{MJD2000.0} &
  \multicolumn{1}{c}{RA} &
  \multicolumn{1}{c}{DEC} &
  \multicolumn{1}{c}{Az} &
  \multicolumn{1}{c}{Elevation} &
  \multicolumn{1}{c}{Distance to Sun} &
  \multicolumn{1}{c}{Distance to Moon} \\ \hline
  \endhead

\hline \multicolumn{7}{r}{{Continued on next page}} \\ \hline
\endfoot

\hline \hline
\endlastfoot

8969.428947 &  33.820 &  -5.245 & 201.515 &  63.458 & 155.0 & 160.7\\
8972.431574 &  34.918 &  -4.038 & 194.802 &  63.052 & 156.5 & 138.4\\
7517.300938 &  58.068 & -42.014 & 303.029 &  33.029 & 118.1 &  90.0\\
7517.352616 &  57.113 & -41.516 & 299.966 &  47.447 & 119.0 &  90.9\\
7517.403542 &  57.004 & -40.727 & 300.549 &  61.370 & 119.6 &  91.0\\
7517.455220 &  56.163 & -40.192 & 315.821 &  75.180 & 120.4 &  91.8\\
7517.506146 &  56.067 & -39.401 &  22.679 &  79.972 & 121.0 &  91.8\\
7517.557824 &  55.248 & -38.860 &  60.143 &  68.177 & 121.9 &  92.5\\
\end{longtable}
\end{center}

\end{document}